# Detecting Data Type Inconsistencies in a Property Graph Database

Joshua R. Porter, Michael N. Young, Aleks Y. M. Ontman

**Abstract –** Some property graph databases do not have a fixed schema, which can result in data type inconsistencies for properties on nodes and relationships, especially when importing data into a running database. Here we present a tool which can rapidly produce a detailed report on every property in the graph. When executed on a large knowledge graph, it allowed us to debug a complex ETL process and enforce 100% data type consistency.

## 1 Introduction

Graph technology is a tool that "forms the foundation of modern data and analytics with capabilities to enhance and improve user collaboration, machine learning models and explainable AI" [1]. Graph databases are organized as nodes connected by relationships, allowing users to efficiently explore connections between data elements. As with relational databases, data is imported into a graph database through an extract/transform/load (ETL) process, which varies depending on the source data, graph database, and desired data model.

There are two main methods for getting data into a graph database: offline and online import. Offline import involves building a database from scratch using files on disk. While this is the fastest method available for importing both nodes and relationships, it has several disadvantages. All data to import must be on disk before building the database, which adds time and complexity if this data must be downloaded from somewhere else. If data transformations are required, this logic must be implemented separately from the data import, potentially adding further time and complexity. A new database can be built this way, but an existing database cannot be updated this way, and users cannot run database queries while the import is happening. Finally, this type of import cannot be made context-aware, i.e., the data cannot be imported differently depending on the state of the database. By contrast, online import of data into a running graph database provides numerous advantages such as allowing context-dependent operations, allowing data to be streamed from another database or an API, maintaining uptime while introducing new data or updating existing data, and potentially simplifying the ETL process. While online import can be slower than offline import due to the overhead of database transactions, this disadvantage can be mitigated significantly with an algorithm that enables parallel processing of relationship data during import, implemented as the iterateRelationship() procedure for Neo4j [2].

One other issue that may occur during online data import is the potential for inconsistent data types. Some graph databases do not rigidly enforce a schema for properties on nodes and relationships. For example, one PHONE node could have a property "phone" whose data type is a string, and another PHONE node could have a property "phone" whose data type is an integer. While in some cases there may be reasons to do this on purpose, it can also happen by accident due to bugs in the ETL code. This can lead to a variety of undesirable results such as duplicate nodes (e.g., two PHONE nodes that represent the same phone number, but have it represented as two different data types), missed connections (e.g., nodes that should be connected by a relationship are not because the query to create the relationship expected a different data type for the node property), and other problems with queries not functioning as expected.

These data type inconsistencies can be difficult to catch, as existing methods for mapping data types (e.g., the apoc.meta.nodeTypeProperties() procedure in Neo4j) consider a limited subset of nodes and relationships. This means that in larger databases, these methods may fail to uncover data type inconsistencies outside of their limited sample. To address this issue, we developed the DeepInspector tool, which is designed to rapidly inspect every property on every node or relationship in the graph to produce a map of properties and their data types, uncovering inconsistencies and allowing developers to fix data type issues at their sources to maintain the integrity of the data.

## 2 Methods

DeepInspector was implemented in Java as two procedures to run on the Neo4j graph database. The procedures, called deepInspect.nodes() and deepInspect.relationships(), were loosely based on the apoc.periodic.iterate() procedure for Neo4j and require the APOC library to be installed. When these procedures are run, they examine every node or relationship in the graph and produce a detailed report in JSON format of the count of each property on each type of node or relationship and their data type(s). These procedures can utilize parallel processing for high performance on large databases.

Each procedure can take the following configuration parameters:

- parallel (default = false): whether to run deep inspection in parallel
- concurrency (default = 50): how many threads to use when running in parallel (requires parallel = true)
- limit (default = 0): specify a limited number of nodes/relationships to inspect for testing purposes. 0 = no limit (inspect the entire database).
- batchSize (default = determined automatically): specify the batch size when nodes/relationships are divided into batches for inspection. This may impact performance.
- debug (default = false): when true, display status messages in the log file to enable an administrator to monitor the progress of the inspection.

## 3 Results

Both deepInspect.nodes() and deepInspect.relationships() were tested with Neo4j Community Edition 4.2.5 and APOC 4.2.0.0 running on an Amazon Web Services EC2 instance (r6g.xlarge, 4 virtual CPUs, 32 GB of memory). All tests were performed on a knowledge graph with 30,086,483 nodes and 585,461,277 relationships. The full graph (~26 GB) was loaded into memory using apoc.warmup.run() before testing. Tests were conducted with parallel set to "false" as a baseline and then with parallel set to "true" for a range of different concurrencies. All tests were performed in triplicate and the results averaged.

Figure 1 shows the time taken to run deepInspect.nodes() for different numbers of concurrencies; Figure 2 shows the same for deepInspect.relationships(). For both nodes and relationships, higher levels of parallelism resulted in improved performance, though gains were marginal beyond a concurrency of 4. This was likely due to limits on the number of threads in Neo4j Community Edition as well as the CPU capacity of the virtual machine. With the highest concurrency settings, DeepInspector generated a detailed report on all the node properties in the graph in 18.6 seconds on average and on all the relationship properties in 117.2 seconds on average. An example section of the report generated is shown in Figure 3.

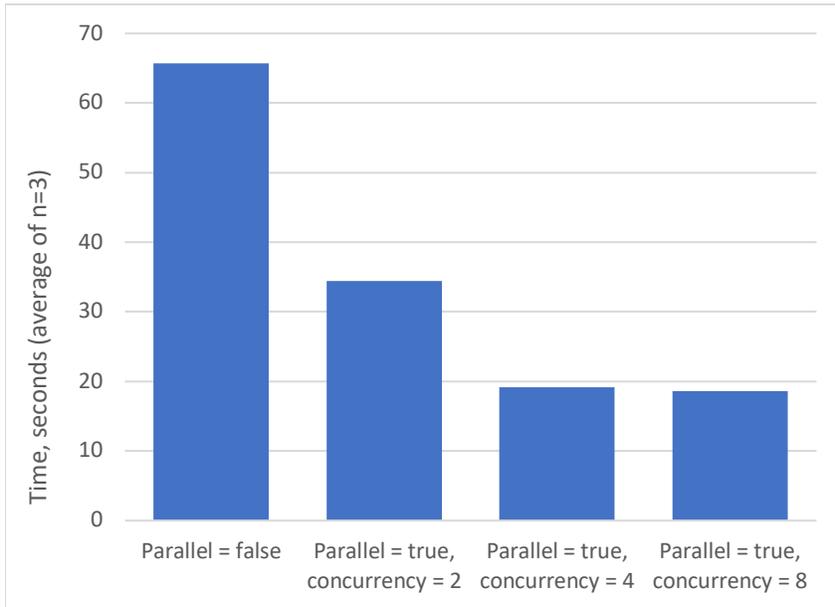

*Figure 1: Time to run deepInspect.nodes() on 30,086,483 nodes of a large knowledge graph using different levels of parallelism and concurrency. All measurements represent the average of n=3 trials.*

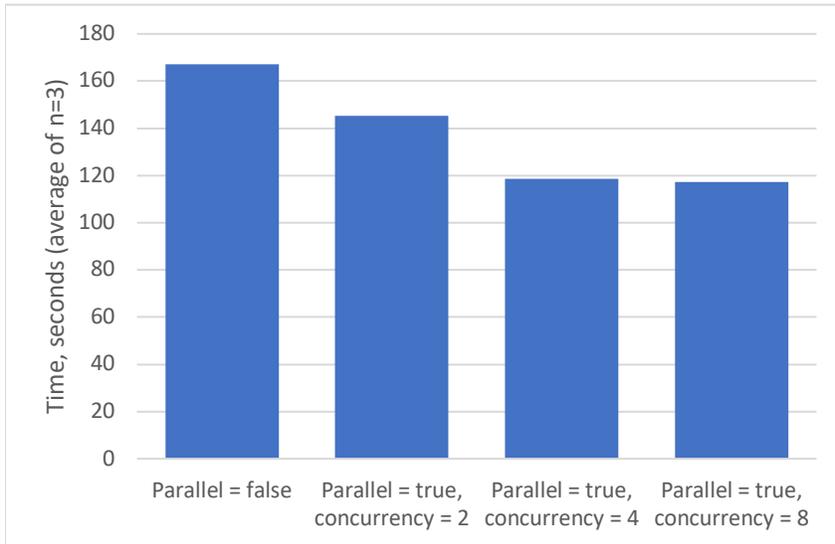

*Figure 2: Time to run deepInspect.relationships() on 585,461,277 relationships of a large knowledge graph using different levels of parallelism and concurrency. All measurements represent the average of n=3 trials.*

```
{
    "PROJECT": {
        "project_start": {
            "LocalDate": 444589
        },
        "total_cost": {
            "Long": 423036
        },
        "project_end": {
            "LocalDate": 444328
        },
        "title": {
            "String": 472445
        },
        "project_num": {
            "String": 488501
        }
    },
    "PERSON": {
        "name": {
            "String": 11962203
        },
        "person_id": {
            "Long": 11967747
        }
    },
…
```

*Figure 3: Example JSON output from running deepInspect.nodes() on a large knowledge graph.*

We used DeepInspector in practice to check data type consistency on a large (> 800 GB) knowledge graph and guide our efforts to troubleshoot the online data import process for this database. As a result, we were able to identify and fix several bugs in the process and guarantee 100% data type consistency afterward, which resolved several issues that had previously frustrated analysts who were using the graph.

## 4 Conclusion

DeepInspector is a versatile tool that produces a detailed high-level view of the data present in a graph database. It can be used to detect developing issues with an ETL process such as missing properties or inconsistent data types. This is invaluable for debugging queries that add or update data, especially in a database that is fed by many such queries. The use of this tool, along with the iterateRelationship() procedure, allows developers to circumvent certain disadvantages of importing data into a running graph database, making online data import a workable method for graph ETL and allowing developers to realize its advantages in simplicity and flexibility.